\documentclass[twocolumn,showpacs]{revtex4}

\usepackage{graphicx}
\usepackage{dcolumn}

\def\thin{\thinspace}

\begin{document}

\title{The glass transition of glycerol in the volume-temperature plane}

\author{Kyaw Zin Win}\email{win@physics.umass.edu} \author{Narayanan Menon}
\email{menon@physics.umass.edu}
\affiliation{
University of Massachusetts, Amherst, MA 01003, U.S.A.
}

\date{\today}

\begin{abstract}
We assess the relative importance of spatial congestion  and lowered
temperature in the slowing dynamics of supercooled glycerol near the
glass transition. We independently vary both volume, $V$,  and
temperature, $T$,  by applying high pressure and monitor the
dynamics by measuring the dielectric susceptibility. Our results
demonstrate that both variables are control variables of comparable
importance. However, a generalization of the concept of fragility of
a glass-former shows that the dynamics are quantitatively more
sensitive to fractional changes in $V$ than $T$. We identify a
connection between the fragility and a recently proposed
density-temperature scaling which indicates that this conclusion
holds for other liquids and polymers.
\end{abstract}


\maketitle

When a liquid is cooled below its melting point its viscosity
increases very rapidly with decreasing temperature until it turns
into a glass, with solid-like properties. For instance, glycerol, a
typical glass-former, shows a remarkable increase in viscosity and
relaxation time of 14 orders of magnitude as it is cooled from 350K
to 180K at 1 atmosphere \cite{menon}. There are two broad classes of
explanations for this rapid increase of viscosity.  One set of ideas
identifies density as the crucial variable, arguing that the
constraints on molecular rearrangements imposed by the dense packing
in a liquid progressively increase due to the thermal contraction
that accompanies cooling, until finally all motions are arrested in
the glassy state. By this argument, the driving force for the rapid
increase of viscosity in the case of glycerol would primarily be the
10\% decrease in volume upon cooling. The second class of
explanations for the slowing dynamics emphasizes the role of
temperature: lowered temperature renders molecules too inactive to
move around and surmount the energy barriers that impede exploration
of their environment.  In order to resolve the issue of whether it
is temperature, $T$, or volume, $V$, that is the dominant variable
in this phenomenon, experiments are required that independently control these variables.
  However, data of this kind are relatively sparse
\cite{roland}.  The objective of the experiments reported in this
Letter is to distinguish unambiguously the effects of
constrained volume and lowered temperature, by using high pressure as a
 mean of independently changing the density of the liquid.

Glycerol is a widely-studied glass-former with broad industrial use.
It is also an experimentally convenient sample: it has a high
dielectric constant and does not easily crystallize. Moreover, some
complementary high pressure data exist for glycerol
\cite{herbst,cook,johari,paluch}. These studies  are principally
isothermal experiments in which temperature is fixed and pressure is
varied in discrete steps; as such, these data are not optimal for
studying temperature dependence. They also appear to disagree with
each other on the temperature dependence of the relaxation frequency
$\nu_p$ near the glass transition temperature, $T_g$, defined as the
temperature at which $\nu_p=0.01$\thin Hz. The temperature
dependence is often characterized by the fragility, 
$m$, defined by $m\equiv-\partial(\log\nu_p)/\partial(T_g/T)|_{T=T_g}$ \cite{angell}.
The fragility thus
  quantifies deviation from the Arrhenius, or
thermally-activated behaviour, and  categorizes the temperature
dependence of glass-formers under isobaric cooling as strong (small
$m$) and fragile (large $m$). One set of high pressure experiments
\cite{cook} finds that the fragility of glycerol increases with high
pressure while another \cite{johari} finds no pressure dependence of
the fragility. These results are not necessarily
contradictory since they are taken in different frequency regimes,
involving different degrees of extrapolation to infer the behaviour
at $T_g$. The present measurements go down to $0.01$\thin Hz,
obviating the need to extrapolate down to $T_g$ in order to infer
fragility. We have also performed an isobaric experiment by
controlling pressure, $P$, and studying the temperature dependence
in detail. We have restricted ourselves to pressures below 1\thin
GPa and taken data at several closely separated pressures to study
changes in dynamics as functions of $P$ and $V$ as well.

\begin{figure}
\includegraphics[width=75mm,height=70mm]{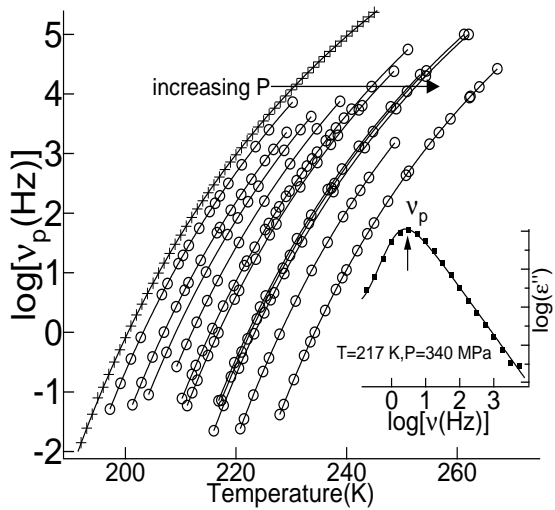}
\caption{\label{fig:cohen}
 Inset:  $\epsilon''$ vs. $\nu$ at 217 K and 340
MPa.  The line is Davidson-Cole fit: $\epsilon=\epsilon_\infty+
(\epsilon_0-\epsilon_\infty)/(1-i\nu/\nu_\alpha)^\beta$.
 The relaxation frequency, $\nu_p$
 is defined to be the peak frequency.
  Main figure: Isobaric plot of  $\nu_p$  vs. $T$.
  The circles represent data at 83, 164, 212, 278, 358, 392, 527, 539, 560, 689, 876 $\pm$ 4 MPa.
  The crosses are 1 atmosphere data \cite{menon}.
 The solid lines are Cohen-Grest fits: $\log\nu_p=A+T-b/[T-T_c+\sqrt{(T-T_c)^2+CT}$.  $A$, $b$, and $C$
 are pressure-independent.  $T_c=V_aP$    where  $V_a$ is a volume parameter.
  The theory predicts that the molecular volume, $V_m=4.605(b/C)V_a$  whence $V_m=1700$\AA$^3$.}
\end{figure}

\begin{figure*}
\includegraphics[width=138mm,height=65mm]{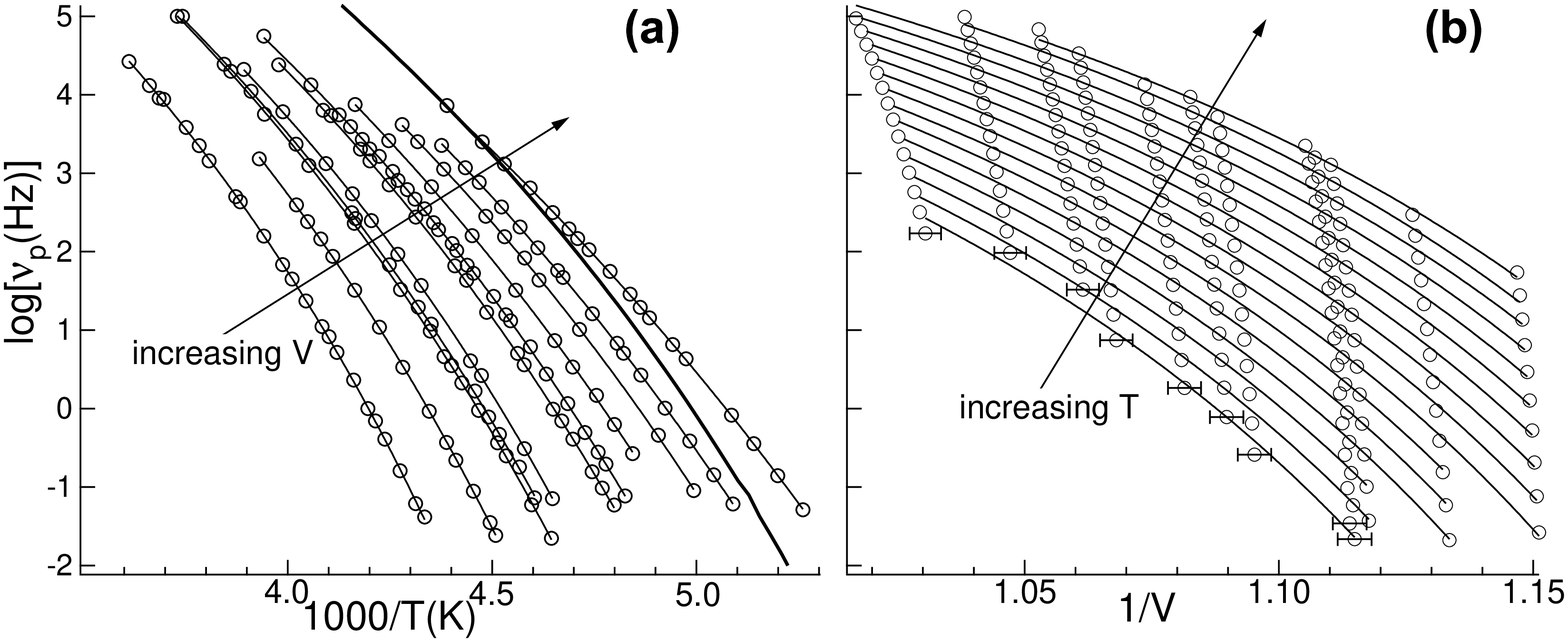}
\caption{\label{fig:isocho}
(a) Isochoric plot.  The solid lines are Vogel-Fulcher-Tammann-Hesse (VFTH) fits:
$\nu_p=\nu_\infty\exp[E/(T_0-T)]$.
The heavy line shows 1 atmosphere isobaric data \cite{menon}.
  $V\equiv 1$ at 273.15\thinspace K, 1\thinspace atmosphere.
  Data are shown for $V$=0.970,0.951,0.942,0.931,0.921,0.916,0.900,0.895,0.895,0.878,0.860 $\pm$ 0.3\%.
(b) Isothermal plot.
  These data are obtained from isothermal cuts to Fig.~\ref{fig:cohen}.  The temperatures from left to right are: {214,216,218,220,222,224,226,228,230,232,234,236,238,240,242\thinspace K}.
The solid lines are Doolittle fits: $\nu_p=\nu_\infty\exp[D/(V_0-V)]$.
}
\end{figure*}

We employ a standard "Teflon cap" technique, widely used in
high-pressure experiments \cite{walker}.  The pressure is inferred
from the resistivity of a manganin coil immersed in the liquid.
Manganin is an alloy whose resistivity as a function of $P$ and $T$
is known \cite{andersson}.  After the liquid is squeezed to a high
pressure at room temperature, we cool it down to low temperatures.
The dynamics of the liquid is studied by measuring the capacitance
of a coaxial capacitor immersed in the liquid.  We obtain the
complex dielectric susceptibility
$\epsilon(\nu)=\epsilon'(\nu)+i\epsilon''(\nu)$  for $\nu$ ranging
from $10^{-2}$ to $10^5$ Hz at each temperature. An example of the
imaginary part of the response, $\epsilon''(\nu)$  is displayed in
the inset of Fig.~\ref{fig:cohen}.  We extract the characteristic relaxation
frequency, $\nu_p$, defined as the frequency of the peak in
$\epsilon''(\nu)$, at each $P$ and $T$. In addition to the
characteristic frequency, $\epsilon(\nu)$ also yields the
distribution of relaxation times and the ionic conductivity \cite{kzw_nm}. In this
Letter we focus on the dependence of $\nu_p$ on $T$, $V$, and $P$.
In Fig.~\ref{fig:cohen} we display $\nu_p$ vs. $T$ for 12 different pressures. As
is evident from the figure, pressure has a strong effect on
relaxation dynamics: our maximum pressure of about 1\thinspace GPa
decreases the relaxation frequency at a fixed temperature by about
seven decades.

Turnbull and Cohen \cite{turnbull} proposed a theory in which the
slowing dynamics in a supercooled liquid is due to reduced free
volume at low temperature.  Free-volume theory predicts that the
viscosity at a fixed temperature is given by  $\eta=\exp(V_m/V_f)$
where $V_m$  is the molecular volume and $V_f$  the free volume. The
temperature dependence of  $\eta$ was computed by Cohen and Grest
\cite{cohen} who derived a formula for the viscosity at a fixed
pressure.  Using the Einstein-Debye relation, $\nu_p=BT/\eta$,
where $B$ is a constant with the dimension of volume, we compare
this prediction to the data in Fig.~\ref{fig:cohen}.  While the fit is good, the
fit parameters lead to a prediction for a mean molecular volume of
$1700$\thin\AA$^{3}$, in excess of length scales from neutron
scattering \cite{dawidowski} or even of a simple estimation of
$123$\thin\AA$^{3}$ from the mass density and molar mass.

A much more direct way to test free-volume theory is to study the
dynamics when the volume -- and therefore the free volume -- is
held fixed. Using an empirical equation of state for glycerol we
calculate \cite{kzw_nm} the volume corresponding to each $T$ and $P$
in Fig.~\ref{fig:cohen}.  This yields the isochoric and isothermal plots shown in
Fig.~\ref{fig:isocho}.  Free-volume theory predicts \cite{turnbull}~that the
viscosity varies only weakly with temperature ($\eta\sim\sqrt{T}$)
at constant volume. Fig.~\ref{fig:isocho}(a) shows that the isochoric change of
$\nu_p$ with respect to temperature is a factor of $10^5$, much
faster than the factor of $1.1$ predicted by the free volume model.

\begin{figure*}
\includegraphics[width=158mm,height=75mm]{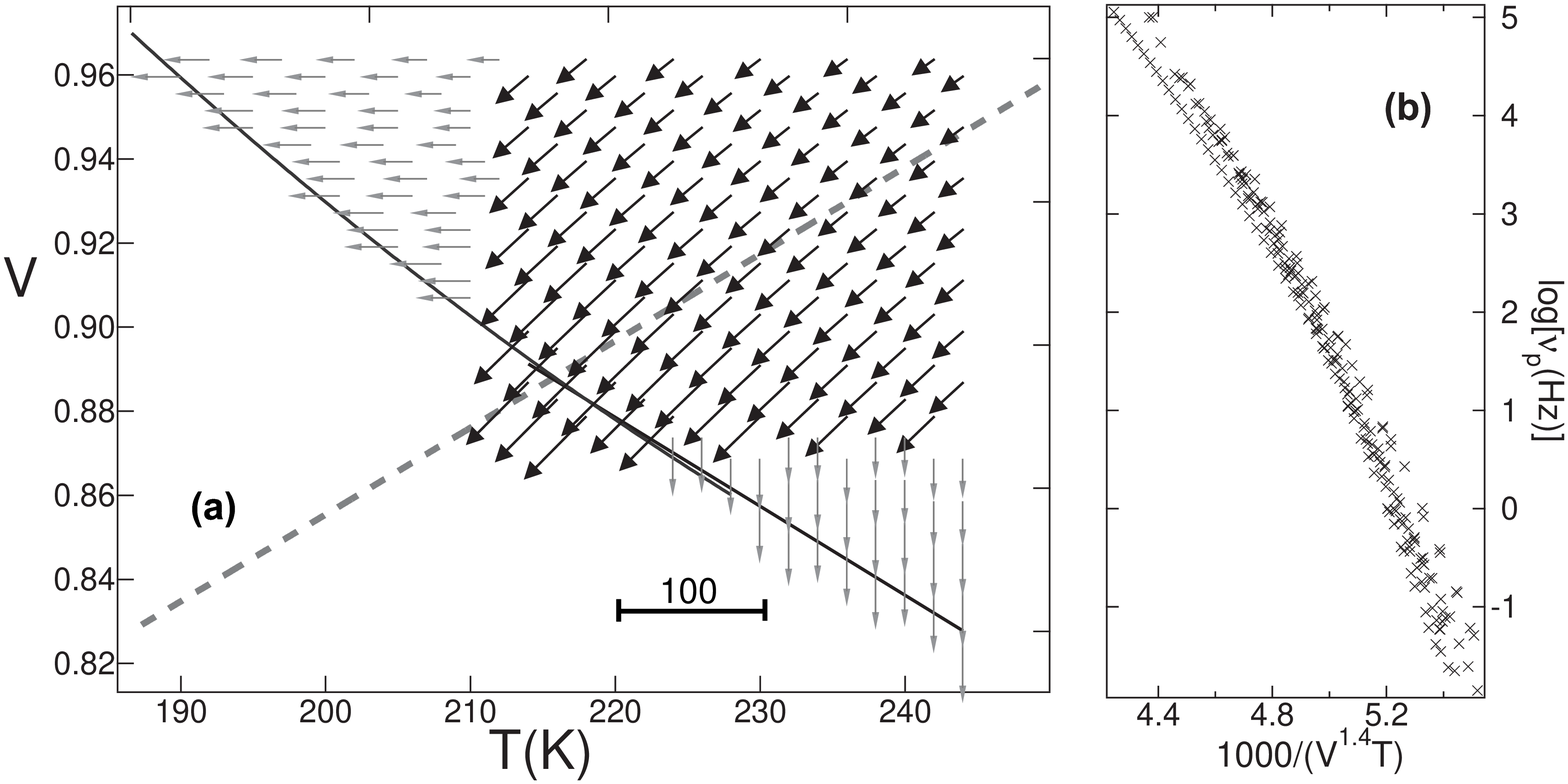}
\caption{\label{fig:vector}
(a) Field of  fragility vector $(M_T,M_V)$.  The scale bar shows the size of the vectors.  The dashed line is a 45$^\circ$ reference line showing that $M_V>M_T$ for all vectors.
  We can evaluate only one component for the grey vectors.  The solid curves from top left to bottom right show the glass transition line of constant relaxation frequency  0.01\thinspace Hz; the horizontal and vertical grey vectors at this glass transition line are the isochoric and isothermal fragilities respectively.  (b) Test of V-T scaling, $\log\nu_p=f(V^\gamma T)$.  The best possible collapse of data is achieved with $\gamma=1.4$}
\end{figure*}

\begin{figure}
\includegraphics[width=82mm,height=58mm]{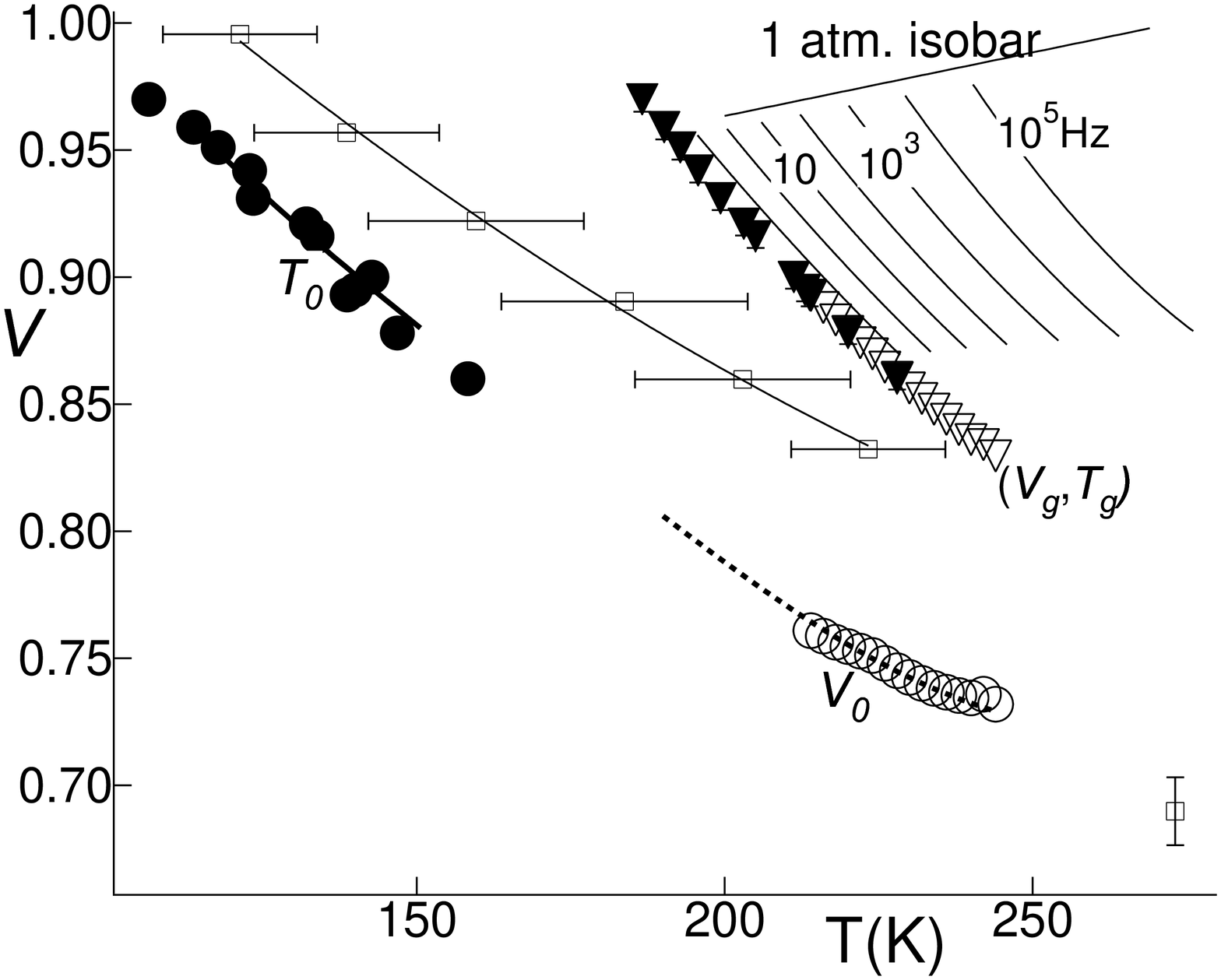}
\caption{\label{fig:master}
 Glass transition in VT plane.  The solid lines are constant frequency contours.
  The error bars are of comparable size to the symbols. The solid circles are $T_0$ from isochoric
VFTH fits and the open circles $V_0$
from isothermal fits to the Doolittle equation to data sets that extend to 0.01\thinspace Hz.
  The open squares are taken from Ref \cite{cook}. The dashed lines involve greater extrapolations of the data.
}
\end{figure}

It is clear from Fig.~\ref{fig:isocho}(a) that even when the volume is constant, the
dynamics remains strongly temperature dependent.   Indeed, we
observe from Fig.~\ref{fig:isocho}(a) that the 1 atmosphere isobaric curve has only a
slightly stronger temperature dependence compared to nearby
isochoric curves.  This means that there is only a small volume
contribution to the slowing dynamics when the liquid is supercooled
along the 1 atmosphere isobar. This observation has led Ferrer {\it et al.}
\cite{ferrer} to conclude that temperature is overwhelmingly more
important than volume. Does this more generally imply that
temperature is the dominant variable in the glass transition and
that the effects of volume are secondary?

To address this question we isolate the role of volume that of temperature by
in an isothermal plot, such as the one shown in Fig.~\ref{fig:isocho}(b) in
which volume is varied, holding $T$ fixed. We observe in Fig.~\ref{fig:isocho}(b)
that $\nu_p$ changes strongly with $1/V$, faster than $\exp(1/V)$.
There appears to be greater variation in $\nu_p$  in the isochoric
cases, however we point out that $1/V$  increases only 14\% in Fig.~\ref{fig:isocho}(b)
 whereas $1/T$ increases by 54\% in Fig.~\ref{fig:isocho}(a).

To assess the relative contribution of $V$ and $T$ it is therefore
important to take into account equivalent fractional changes in
these variables. To this end we compare $\Delta(\log\nu_p)|_V$  for
some  $\Delta T/T$ with $\Delta(\log\nu_p)|_T$  for the same $\Delta
T/T$. This leads us to consider the variables,
$M_T\equiv-T(\partial/\partial T)\log\nu_p|_V$ and
$M_V\equiv-V(\partial/\partial V)\log\nu_p|_T$.  When evaluated at
$\nu_p=0.01$ Hz, $M_V$  is the isothermal fragility with respect to
$V$ and $M_T$  the isochoric fragility with respect to $T$.  We call
these quantities the generalized fragilities by analogy to the
traditional fragility, $m$, which   compares glass formers with
different  $T_g$ because $m$ measures the rate of slowing dynamics
only after scaling out the material-dependent $T_g$.  Likewise $M_V$
and $M_T$ compares the rate of change of $\nu_p$   in $V$ and $T$
directions only after scaling out the relevant variables.
Furthermore, the variables $M_V$  and $M_T$ are manifestly symmetric
in $V$ and $T$.

We display in Fig.~\ref{fig:vector}(a) the vector field  $(M_T,M_V)$ plotted in the
V-T plane. At all places where we can evaluate $M_T$  and $M_V$,
they are comparable: $1.2<(M_V/M_T)<1.7$.  Thus the effects on
relaxation frequency of equal fractional changes in volume or
temperature are comparable over the entire experimental range, and
neither variable can be neglected. While neither variable is
overwhelmingly dominant, it is clear that $M_V>M_T$  over the
experimentally accessible region. Thus relaxation frequencies are
more sensitive to fractional changes in  $V$ than $T$ everywhere in
the V-T plane.  In particular, the data indicate that this is also
true for the 1 atmosphere isobaric glass transition at
$T_g=192$\thinspace K, $V_g=0.959$. While we can only directly
measure $M_T$  at (192,0.959) a linear extrapolation of $M_V$ to the
smallest temperature in Fig.~\ref{fig:vector}(a) shows that  $M_V/M_T=1.58$.

There are two opposing factors affecting the fragilities as one
moves along the glass line towards high densities: increasing $T$
reduces fragility while decreasing $V$ makes it bigger. It is clear
from Fig.~\ref{fig:vector}(a) that the latter effect dominates, inasmuch isochoric and
isothermal fragility both increase along the glass transition line
towards lower $T$ or lower $V$. The increase of isochoric fragility,
$M_T$, with density is an additional indicator of greater
sensitivity to changes in $V$ than $T$. We have observed that the
width of the relaxation too, increases more rapidly in the $V$
direction than the $T$ direction \cite{kzw_nm}.

Are these results specific to glycerol?  We address this issue by
means of a density-temperature scaling suggested by Casalini and
Roland \cite{casalini} and also Alba-Simionesco {\it et al.} \cite{alba}.
For several liquids and polymers they found that
$\log\nu_p=f(V^\gamma T)$  where $\gamma$  and the function $f$ are
material dependent.  Next, we observe that this scaling exponent is
equal to the ratio of the generalized fragilities: $\gamma=M_V/M_T$.
This scaling for our data is shown in the  Fig.~\ref{fig:vector}(b)
with $\gamma=1.4$; this is less than 1.8 obtained using the data of Cook {\it et al.}
\cite{cook} by Ref. \cite{dreyfus}. The scaling in Fig.~\ref{fig:vector}(b)
is by no means perfect, reflecting the fact that $M_V/M_T$ is not
constant. For a power-law potential, $V(r)\sim r^{-3\gamma}$, the
scaling is exact. However, in the more complicated situation of a
molecular liquid, one should not be surprised if the scaling is
inexact because changes in density -- and consequently the
intermolecular spacing -- will emphasize different parts of the
potential.   Nevertheless, to the extent that this scaling is even
approximately successful over a limited range of data, $\gamma>1$
implies that $M_V>M_T$.  There are 25 liquids with  $\gamma>1$ and only one (sorbitol) with
$\gamma<1$ \cite{roland}. Therefore $\gamma>1$  is observed for a wide range of
liquids. The greater sensitivity of relaxation dynamics to
fractional volume changes is thus not limited to glycerol, and
appears to be typical rather than exceptional.

We summarize the new data for glycerol in Fig.~\ref{fig:master} as a set of
constant relaxation frequency curves in the V-T plane.  The glass
transition line ($\nu_p=$0.01 Hz) is formed by $T_g$  and $V_g$
extracted from Fig.~\ref{fig:isocho}.  The constant frequency curves get denser
towards lower frequencies.  Apart from the constant frequency lines
in the experimentally accessible regime we also indicate in Fig.~\ref{fig:master}
extrapolations to the zero frequency limit.  The solid circles
represent $T_0$ from VFTH fits to isochoric data and the open
circles $V_0$ from Doolittle fits to isothermal data.  While the
existence of a finite temperature or finite volume structural arrest
depend solely on extrapolations, it is worth pointing out that these
two independent extrapolations support each other in that they
appear to produce a continuous  $(V_0,T_0)$ curve.  With the available data the value of $T_0$ appears to be
increasing in an unbounded way at high densities.  This agrees with
the expectation that an extremely compressed system will resemble a hard sphere liquid, where $T_0\rightarrow\infty$.  This limit is approached along a
concave curve, as in attractive colloidal systems \cite{trappe}.
Clearly, even broader explorations in the V-T plane of both dynamics
and of the Kauzmann entropy crisis would be useful in assessing the
role of $T_0$ and $V_0$.  We end by recapitulating our principal result: neither axis on
this plane may be neglected in understanding supercooled glycerol.

We acknowledge support from the NSF-DMR $0305396$ and the MRSEC at
UMass. We thank R. Krotkov, S. Brown, I. R. Walker, A. Husmann, and
J. Hu for conversations regarding experimental issues and W. Pollard
for his skilled machining.  Manganin was generously supplied by
Isabellenh\"{u}tte, Germany, via their U.S. distributor, Isotek.
N.M. gratefully acknowledges several inspiring discussions with D.
Kivelson. K.Z.W. thanks C. M. Roland and R. Casalini for useful conversations.

\end{document}